\documentclass[RNAAS]{aastex63}

\newcommand{\UCB}{Department of Astronomy,  University of California Berkeley, Berkeley, CA}

\newcommand{\NIJ}{Department of Astrophysics/IMAPP,Radboud University, Nijmegen, Netherlands}

\newcommand{\USQ}{University of Southern Queensland, Toowoomba, Australia}
\newcommand{\SETI}{SETI Institute, Mountain View, California}
\newcommand{\KZA}{University of Malta, Institute of Space Sciences and Astronomy}

\newcommand{\BPF}{The Breakthrough Initiatives, NASA Research Park, Bld. 18, Moffett Field, CA}

\newcommand{\PENN}{Department of Astronomy and Astrophysics, Pennsylvania State University, University Park, PA}
\newcommand{\MIT}{Massachusetts Institute of Technology, Cambridge, MA}

\newcommand{\Curtin}{International Centre for Radio Astronomy Research, Curtin Institute of Radio Astronomy, Curtin University, Perth, Australia}

\begin{document}

\title{Re-Analysis of Breakthrough Listen Observations of FRB\, 121102: Polarization Properties of Eight New Spectrally Narrow Bursts}

\correspondingauthor{Jakob T. Faber}
\email{jfaber@oberlin.edu}

\author[0000-0001-9855-5781]{Jakob T. Faber}
\affiliation{Department of Physics and Astronomy, Oberlin College, Oberlin, OH}
\affiliation{\UCB}

\author[0000-0002-8604-106X]{Vishal Gajjar}
\affiliation{\UCB}

\author[0000-0003-2828-7720]{Andrew P. V. Siemion}
\affiliation{\UCB}
\affiliation{\NIJ}
\affiliation{\SETI}
\affiliation{\KZA}

\author[0000-0003-4823-129X]{Steve Croft}
\affiliation{\UCB}
\affiliation{\SETI}

\author[0000-0002-8071-6011]{Daniel Czech}
\affiliation{\UCB}

\author[0000-0003-3197-2294]{David DeBoer}
\affiliation{\UCB}

\author{Julia DeMarines}
\affiliation{\UCB}

\author{Jamie Drew}
\affiliation{\BPF}

\author[0000-0002-0531-1073]{Howard Isaacson}
\affiliation{\UCB}
\affiliation{\USQ}

\author[0000-0003-1515-4857]{Brian C. Lacki}
\affiliation{\UCB}

\author{Matt Lebofsky}
\affiliation{\UCB}

\author{David H.\ E.\ MacMahon}
\affiliation{\UCB}

\author{Cherry Ng}
\affiliation{\UCB}

\author{Imke de Pater}
\affiliation{\UCB}

\author[0000-0003-2783-1608]{Danny C.\ Price}
\affiliation{\UCB}
\affiliation{\Curtin}

\author{Sofia Z. Sheikh}
\affiliation{\PENN}

\author{Claire Webb}
\affiliation{\MIT}

\author{S. Pete Worden}
\affiliation{\BPF}

%



\begin{abstract}
    We report polarization properties for eight narrowband bursts from FRB \,121102 that have been re-detected in a high-frequency (4--8\,GHz) Breakthrough Listen observation with the Green Bank Telescope, originally taken on 2017 August 26. The bursts were found to exhibit nearly $100\%$ linear polarization, Faraday rotation measures (RM) bordering 9.3$\times$10$^4$\,rad-m$^{-2}$, and stable polarization position angles (PA), all of which agree with burst properties previously reported for FRB \,121102 at the same epoch. We confirm that these detections are indeed physical bursts with limited spectral occupancies and further support the use of sub-banded search techniques in FRB detection.
\end{abstract}

\section{Introduction}
Fast Radio Bursts (FRBs) are a class of highly luminous, heavily dispersed, millisecond-duration radio transient events \citep{Lorimer2007,Thornton2013,Petroff2016}. 
FRB\, 121102 is one of the first FRBs to emit repeatedly \citep{Spitler2014,Spitler2016} and has been successfully localized to a dwarf galaxy at $z = 0.19273(8)$ \citep{Tendulkar2017, Chatterjee2017}. 
\cite{Gajjar2018} reported a high-frequency (4--8\,GHz) detection of 21 bursts from this source with the Robert C. Byrd Green Bank Telescope (GBT), conducting a dispersion search across the full 4\,GHz band. Shortly thereafter, \cite{Zhang2018} devised a machine-learning-based method of transient searching with convolutional neural networks that uncovered 72 additional low-energy, narrowband bursts. Until now, polarization properties have not been reported for these new pulses. Here, we perform a sub-banded dispersion search \citep{Foster2018,Kumar2020} of these datasets, and report a re-detection of 20 bursts first discovered by \cite{Zhang2018}. We present detailed polarization properties for eight bursts with sufficiently high signal-to-noise (S/N) ratios and find their rotation measures (RM) and polarization position angles (PA) to be consistent with previously reported  
measurements of bursts detected at the same epoch \citep{Gajjar2018, Michilli2018}. Our results suggest that these weaker bursts with limited spectral coverage ($\leq$ 500\,MHz) are real and exhibit similar polarization properties to broadband bursts.  





\section{Observations and analysis} 
Breakthrough Listen (BL) is conducting one of the most comprehensive surveys for evidence of intelligent life in the Universe \citep{Worden2017}. The GBT is one of the primary observing facilities of the BL program, where a state-of-the-art 64-node GPU cluster has been deployed \citep{MacMahon2018}. FRB\, 121102 was observed with the BL backend on UT 2017 August 26 during a 5-hour BL observing block with ten 30-minute scans. Flux and polarization calibration scans were performed using an observation of 3C 161 and an off-source position, alongside the calibration noise diode, for one minute each. A test pulsar, PSR\, B0329+54, was also observed to validate the flux and polarization calibration pipeline. 

The BL backend at the GBT enables the storage of 8-bit baseband complex raw voltages at the native resolution of 0.34\,$\mu$sec \citep{MacMahon2018}. The data presented here were recorded across 4--8\,GHz between 32 compute nodes, each node recording an overlapping 187.5\,MHz of sub-band. The use of baseband data offers significant computational advantages for studying FRBs and other dynamic radio transients, since reduced data products are often prone to having their polarization properties masked by channelization. This masking can lead to both improperly measured and corrected RM in sources with extreme local conditions, such as FRB\, 121102 \citep{Michilli2018}. All the raw baseband voltage data recorded from these observations are publicly available\footnote{http://seti.berkeley.edu/blog/frb-data}. 

The baseband data from 10 scans were first reduced to total intensity SIGPROC formatted filterbank files \citep{Lebofsky2019} which are also publicly available\footnote{https://seti.berkeley.edu/frb-machine/overview.html}. \cite{Gajjar2018} reported 21 bursts using a locally developed, Heimdall-based \citep{Barsdell2012} single-pulse detection pipeline named SPANDAK\footnote{https://github.com/gajjarv/PulsarSearch}. Subsequently, \cite{Zhang2018} reported 72 new spectrally limited bursts using a novel CNN based ML technique. The substantial increase in detection rate between \cite{Gajjar2018} and \cite{Zhang2018} made it clear that methods which search the larger bandwidth (in our case, 4\,GHz) are largely insufficient for resolving bursts with low spectral occupancies. Hence, we divided the band into 8 sub-bands (each spanning $\sim$ 500\,MHz) and searched each of them using the SPANDAK pipeline. In order to identify bursts which exceeded the 500\,MHz sub-bandwidth, detections were cross-referenced between sub-bands to identify those coincident within one dispersion time delay across the full 4\,GHz band ($\sim$ 0.11\,s). In total, the search re-detected all bursts reported by \cite{Gajjar2018}, as well as an additional 20 detected by \cite{Zhang2018}, 8 of which exhibited S/N ratios sufficient for calibration and polarimetry analysis. 

\section{Results}
All eight bursts reported here were detected within the first 30 minutes of a 5-hour observation. The majority of bursts found by \cite{Gajjar2018} also fell within the same detection window, indicating a higher source activity early on that waned over the course of the observation. We extracted 0.6\,seconds of baseband voltages around the detection of these bursts and carried out polarization calibration procedures similar to those discussed by \cite{Gajjar2018}. 

The eight spectrally narrow bursts presented in Table \ref{table:1}, for which polarization spectra are publicly available\footnote{http://seti.berkeley.edu/papers/FRB\_121102\_RNAAS\_PA\_Plots.pdf}, exhibit nearly 100\% linear polarization with a stable PA and extremely high RM values of 9.3$\times$10$^4$\,rad-m$^{-2}$. We find our measurements are consistent with those reported for FRB\, 121102 \citep{Gajjar2018,Michilli2018}. While the narrowband structure may have called into question the physical validity of these bursts \citep{cs19}, or suggested an alternate emission mechanism compared to the type that produces conventional broadband bursts, we find the bursts reported in Table \ref{table:1} are indeed just low-energy bursts from FRB\, 121102 with limited spectral occupancies. \cite{Gourdji2019} also reported a nontrivial number of low-energy bursts from FRB\, 121102 with comparably limited spectral occupancies and suggested they originated from propagation effects such as plasma lensing as opposed to variability of the engine. While we cannot confirm this directly, our success in uncovering undetected narrowband bursts with polarization properties similar to bursts with greater spectral occupancies is in agreement with \cite{Gourdji2019}. Thus, our findings strongly encourage the implementation of sub-banded searching for all FRBs---repeaters in particular, due to their complex and spectrally limited morphologies \citep{Foster2018, Kumar2020}. Finally, we confirm the physical nature of these low-energy, narrowband bursts which further bolsters the suggested highly magneto-ionic nature of the source's host environment, as was first posited by \cite{Michilli2018} and \cite{Gajjar2018}.

\tabcolsep=0.09cm
\begin{table}[ht]
\caption{\textbf{Properties of low-energy spectrally narrow bursts from FRB\, 121102 at C-Band.} Burst arrival time; MJD referenced to infinite frequency at the solar system barycenter; DM for maximum  S/N; RMs in the observer's frame; mean PAs referenced to infinite frequency; calibrated fluence; burst temporal width; burst spectral width.
}              
\label{table:1}      
\centering                                      
\begin{tabular}{c c c c c c c c c}          
\hline\hline                        
{\scriptsize Burst} & {\scriptsize TOA (s)} & {\scriptsize MJD (57991+)} & {\scriptsize {\ $\mathrm{DM}_{\mathrm{S} / \mathrm{N}}\left(\mathrm{pc }~\mathrm{cm}^{-3}\right)$}} & {\scriptsize $\mathrm{RM}_{\mathrm{obs}}\left(\mathrm{rad} ~\mathrm{m}^{-2}\right)$} & {\scriptsize $\mathrm{PA}_{\infty}^{\mathrm{mean}}(\mathrm{deg})$} & {\scriptsize Fluence (Jy $\mu$s)} & {\scriptsize Width (ms)} & {\scriptsize Spectral Width (MHz)}\\
\hline   
    1  & 26.404  & 0.410021826 & $646.1$ & $93690(55)$  & $74(2)$  & $84(2)$   & $0.87(1) $ & $220$ \\ 
    2  & 263.405 & 0.412764885 & $631.4$ & $93650(25)$  & $64(3)$  & $69(3)$   & $2.13(9) $ & $640$ \\ 
    3  & 277.366 & 0.412926475 & $636.6$ & $93736(107)$  & $58(2)$  & $94(2)$   & $2.50(2) $ & $650$\\ 
    4  & 315.041 & 0.413362526 & $589.3$ & $94039(56)$  & $74(3)$  & $65(2)$   & $2.23(3) $ & $280$\\ 
    5  & 558.922 & 0.416185231  & $527.0$ & $93910(30)$  & $65(2)$  & $91(2)$   & $1.28(4) $ & $220$\\ 
    6  & 652.592 & 0.417269372  & $578.0$ & $93430(90)$  & $66(3)$  & $46(2)$   & $0.67(1) $ & $600$\\ 
    7 & 1071.803 & 0.422121351  & $595.8$ & $93655(25)$  & $67(2)$  &$ 101(2)$  & $1.40(7) $ & $200$\\ 
    8 & 1440.944 & 0.426393824  & $622.9$ & $93556(50)$   & $71(3)$  & $124(3)$  & $1.76(4) $ & $120$\\ 
    
\hline                                             
\end{tabular}
\end{table}

\acknowledgments

Breakthrough Listen is managed by the Breakthrough Initiatives, sponsored by the Breakthrough Prize Foundation. The Green Bank Observatory is a facility of the National Science Foundation, operated under cooperative agreement by Associated Universities, Inc. Jakob Faber and Steve Croft were supported by the National Science Foundation under the Berkeley SETI Research Center REU Site Grant No. 1950897.

\bibliography{rnaas}{}
\bibliographystyle{aasjournal}

%
%
%

\end{document}